\documentclass[10pt,letterpaper,twocolumn]{article} 

\usepackage{ol2}
\usepackage{amsmath}
\usepackage{epstopdf}
\usepackage{url}

\begin{document}

\twocolumn[ 

\title{Invariance of waveguide grating mirrors\\ to lateral displacement phase shifts}

\author{Daniel Brown$^{1,*}$, Daniel Friedrich$^{2,**}$, Frank Br\"{u}ckner$^{1}$,\\ Ludovico Carbone$^{1}$, Roman Schnabel$^{2}$ and Andreas Freise$^{1}$}

\address{
$^1$University of Birmingham, School of Physics and Astronomy, Edgbaston, Birmingham, B15 2TT, UK
\\
$^2$Institut f\"ur Gravitationsphysik, Leibniz Universit\"{a}t Hannover and Max-Planck-Institut f\"{u}r Gravitationsphysik (Albert-Einstein-Institut), Callinstrasse 38, 30167 Hannover, Germany
\\
$^*$Corresponding author: ddb@star.sr.bham.ac.uk \\
$^{**}$Currently working at the Institute for Cosmic Ray Research, University of Tokyo, Japan
}

\begin{abstract}
We present a method to analyse the coupling of lateral displacements in nanoscale structures, in
particular waveguide grating mirrors (WGM), into the phase of a
reflected Gaussian beam using a {\it finite-difference time-domain}
simulation. Such phase noise is of interest for using WGMs in 
high-precision interferometry. We show that, to the precision of our simulations ($10^{-7}$rad), waveguide mirrors do not couple lateral displacement into phase noise of a reflected beam and that WGMs are therefore not subject to the same stringent
alignment requirements as previously proposed layouts using diffraction gratings.
\end{abstract}


 ] 

\noindent The sensitivity of high precision interferometry experiments
such as the preparation of entangled test masses \cite{PhysRevLett.100.013601}, frequency stabilisation with rigid cavities
\cite{PhysRevLett.93.250602} and gravitational wave detectors
\cite{0264-9381-19-5-305}
are eventually limited by the quantum noise of the interrogating light
or the thermal noise of the optical components. Quantum shot noise of the
detected light can typically be reduced by increasing the laser power. However, this
induces larger thermal distortions in 
the high-reflectivity (HR) coatings and substrates.
A number of new techniques have been suggested to reduce the
distortions from high power beams as well as thermal noise: the use of non-fundamental beam shapes \cite{Chelkowski09}, all-reflective
interferometer layouts using dielectric gratings to reduce absorption of the laser in
optics \cite{Sun:98} and the use of waveguide grating mirrors
(WGM) \cite{Bunkowski06b}. WGMs would replace HR mirror 
coatings, reducing their thickness by a factor of 10 to 100; which promises to reduce the Brownian coating thermal noise
\cite{0264-9381-19-5-305,Brueckner08}. However, gratings couple lateral displacements into
the phase of diffraction orders $|m|>0$ \cite{Wise05, deepaper}. This
places stringent requirements on the alignment and stability of
gratings and the incident laser beams for high-precision interferometry \cite{Freise07}. WGMs rely on diffraction into the first
order and could potentially be subject to the same displacement phase noise effects. However, so far no theoretical or experimental evidence has been presented to demonstrate whether WGMs suffer from this.

In this Letter we apply a rigorous {\itshape Finite-Difference
    Time-Domain} (FDTD) based simulation with Gaussian beams to show
that WGMs do not couple lateral displacements into the phase of a
reflected laser beam. We further provide a simplified ray picture
to illustrate this result.

\begin{figure}[!Ht]
\centerline{\includegraphics[height=3.25cm]{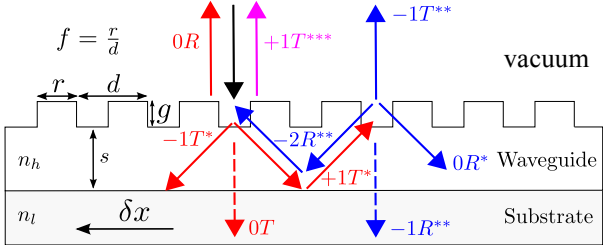}}
\caption{{\small The incident beam (black) is coupled into the waveguide layer by the grating
  into $m=\pm1$. Orders $\pm1T$ propagate along the waveguide coupling back into the
  vacuum (blue) to interfere with the initially reflected light, $0R$,
	picking up $\Delta\Phi_m$ phase terms with each interaction with the grating. 
	Further coupling into the vacuum is also possible (magneta) which involved further
	$\Delta\Phi_m$ terms.
	The $\ast$ superscript refers to the number of diffractions beam has undergone.
	$R=$\,Reflection from grating, $T=$\,Transmission from grating, first number is order of diffraction $m$.}}
\label{fig:waveguide_ray}
\end{figure}

Ray pictures have already been used to describe several WGM features
\cite{Sharon:97} as depicted in Fig.\ref{fig:waveguide_ray}.
WGMs in their most simplistic form consist
of 2 layers: ({\itshape i}) a waveguide layer applied to some substrate material; and ({\itshape ii}) a
grating layer which couples the incident laser light into the
waveguide layer (typically etched into the waveguide
layer). In this case both the waveguide and grating layers are made from a high
refractive index $n_{h}$ material with the substrate material's lower
refractive index denoted by $n_{l}$. The geometrical grating parameters must be
carefully chosen to reach a theoretical maximum reflectivity \cite{Bunkowski06b, Brueckner08}. For given materials and laser light wavelength, the grating period $d$ is chosen such that the normally incident beam is diffracted into the 1$^{\text{st}}$ order within
the waveguide layer at an angle that allows total-internal-reflection
(TIR) at the waveguide-substrate boundary to occur. TIR at the substrate boundary along with the grating create a waveguide in which the $\pm$1$^{\text{st}}$ orders propagate. These undergo diffraction at the grating multiple times, coupling out into the vacuum where it interferes with the reflected specular laser light. The remaining grating parameters, namely the thickness of the waveguide layer
$s$, fill-factor $f$ and groove depth $g$ can all
be tuned to provide destructive interference in the substrate and
constructive in the vacuum, ideally providing 100\%
reflectivity. 

A lateral displacement $\delta x$ of some grating structure versus the incident beam
induces a phase shift of
\begin{equation}
\Delta\Phi_{m} = 2\pi m \delta x/d \label{eq:phaseshift}
\end{equation}
relative to a non-displaced beam for diffraction order $m$ \cite{Freise07}.
For WGMs we require that any rays coupling out into the
vacuum do not have any phase terms dependent on $\delta x$. From Fig.\ref{fig:waveguide_ray} each time a ray is diffracted and picks up a $\Delta\Phi_{m}$ term, an $\ast$ is added as a superscript. The ray $-1T^{\ast\ast}$ diffracted into the vacuum has collected two $\Delta\Phi_{m}$ terms, its total phase is then
$\Phi_{-1T^{\ast\ast}} = \Phi_o(s,d,g,f,n_l,n_h) + \Delta\Phi_{+1} + \Delta\Phi_{-1}$,
where $\Phi_o$ is a collection of all phase terms depending on the WGM
parameters, but not $\delta x$. $\Phi_o$ is tuned with simulations by adjusting each parameter to produce 100\% reflectivity. From Eq.(\ref{eq:phaseshift}) we see the
$\Delta\Phi_{-1}$ and $\Delta\Phi_{+1}$ terms cancel, with a similar
argument being valid for all other rays that couple out into the
vacuum such as $+1T^{\ast\ast\ast}$. Thus, following this
strongly simplified picture any of the phase noise effects outlined in Ref.\cite{Freise07} for gratings should not apply to WGMs under normal
incidence.

In order to provide a rigorous and physically correct computer model of a
finite beam reflected from a WGM we have implemented a numerical
finite-difference time-domain (FDTD) based algorithm, which 
provides the ability to model a variety of grating structures as
well as arbitrary and finite incident electromagnetic field
distributions. The simulation tool is coded in {\itshape Java}, open sourced (\url{http://kvasir.sr.bham.ac.uk/redmine/projects/
fdtd}) and was based on Ref.\cite{taflove:2005}.
A 2D FDTD simulation sufficed for our needs as
only a displacement of the WGM in one direction orthogonal to a normally
incident Hermite-Gaussian (HG) beam was required; thus speeding up computation time
significantly. Two extra
features were also required for the simulation \cite{taflove:2005}:
{\itshape Total-Field Scattered-Field} (TFSF) for separating the
incident and reflected beam from the WGM and {\itshape complex
perfectly matched layers} (CPML) to reduce reflections from the
simulation boundaries. The simulation package was
validated by reproducing known dependencies (found in Ref.\cite{Bunkowski06b}) of the reflectivity as a function of the grating parameters and by
investigating the phase noise of standard diffraction gratings \cite{deepaper}.

\begin{figure}[!Ht]
\centerline{\includegraphics[height=3.5cm]{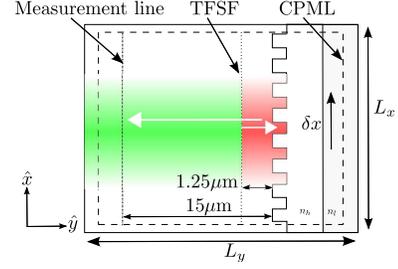}}
\caption{{ \small Schematic layout of 2D FDTD simulation for testing WGM shift
  invariance. Gaussian beam injected along TFSF boundary onto WGM
  (Red). The reflected beam (Green) then propagates to the measurement
  line where the phase is measured. The CPML absorbs outgoing waves to reduce reflections from boundaries.}}
\label{fig:waveguide_fdtd_layout}
\end{figure}

The aim of the simulation was to measure the wavefront of a HG beam
reflected from a WGM whilst displacing it from $\delta x = 0
\rightarrow d$. Along the wavefront the phase can then be deduced and
plotted against $\delta x$ to view any apparent phase shifts. The
simulation setup is depicted in Fig.\ref{fig:waveguide_fdtd_layout},
where a HG TEM$_{00}$ is injected in the $\hat{x}$
direction along the TFSF boundary and the electric field of the
reflected beam is measured along the measurement line $15\mu m$ away
to avoid near-field variations. The {\itshape Courrant stability
  factor} \cite{taflove:2005} for the simulation was chosen as
$S=c\Delta t /\Delta x= 1/\sqrt{2}$; where $\Delta t$ is the
simulation timestep and $\Delta x = \Delta y = 25$nm are the size of
the 2D discretisation of the simulation space with dimensions
$L_y=250\Delta y$ and $L_x=4000\Delta x$. The injected beam had a
wavelength $\lambda=1064$nm and was positioned such that the waist was at the
WGM with size $w_0=800\Delta x=20\mu$m. The WGM parameters chosen were
$d=28\Delta x=700$nm, $g=14\Delta x=350$nm, $f=0.5$ and $s=5\Delta x=125$nm
which provided a reflectivity of 99.8\% for the incident beam (in agreement with Ref.\cite{Bunkowski06b}). The indices of refraction used were fused silica for $n_l= 1.45$
and Ta$_2$O$_5$ for $n_h = 2.084$ which are the typical materials used
for $1064$nm optics.

\begin{figure}[!Ht]
\centerline{\includegraphics[width=0.5\textwidth]{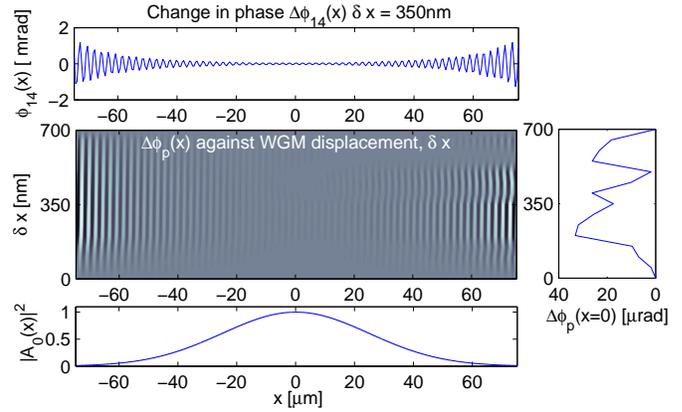}}
\caption{{\small Central plot shows phase change in the reflected beams
  wavefront $\Delta\phi_p(x)$ against WGM displacement, $\delta x$. Top plot shows
  cross section of phase at $\delta x=350$nm where the phase
  change is maximum. The right plot shows the variation in the
  phase of at $x=0$ against $\delta x$. Bottom
  plot shows the intensity of the reflected beam along $x$.}}
\label{fig:results}
\end{figure}

\begin{figure}[!Ht]
\centerline{\includegraphics[width=0.5\textwidth]{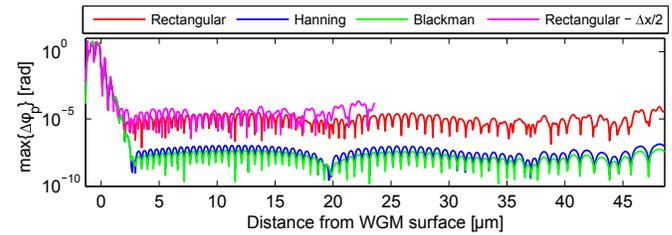}}
\caption{{\small Maximum change in phase at beam peak intensity as the WGM is displaced
  from $\delta x=0\rightarrow d$ with increasing distance from the WGM
  surface and different FFT windowing functions. Large near field phase shifts are seen close to the WGM and
  then a flat noise. Increasing the spatial and temporal resolution of FDTD $\Delta x\rightarrow\Delta x/2$ did not appear to offer
  reductions in numerical noise caused by FDTD.}}
\label{fig:dhpi_disp}
\end{figure}

Eq.\ref{eq:phaseshift} states the phase shift for $m=\pm1$ is periodic with displacements of the grating period $d=28\Delta x$.
Thus for this effect to be visible the simulation was run 28 times for
offsets $\delta x = p\Delta x$ with $p=0,1,2,...,28$. Approximately
3000 timesteps were required for the reflected beam to reach an
approximate steady-state in each simulation. At this point 1024 time
samples of the electric field at each point along the measurement line
were taken, $E_{p}(x, t)$. The generalised Goertzel algorithm (based on Fast Fourier Transforms (FFT)) \cite{springerlink:10.1186/1687-6180-2012-56} was used to extract the
amplitude $A_p(x)$ and phase $\phi_p(x)$ of the reflected beam along the measurement line for the incident laser frequency $f_0=c/1064$nm. $\phi_p(x)$ was obtained for each
offset $\delta x = p\Delta x$ of the WGM with the change in phase with
displacement defined as
$\Delta\phi_p(x)=\phi_p(x)-\phi_0(x)$.
Our model showed that displacement phase shift for WGMs are at least $10^{5}$
smaller than for an equivalent grating setup, see Fig.\ref{fig:results}:
The central plot shows the phase change as a function of the
displacement along the beam profile; the satellite plots provide the
scale for the central plot. 
The top plot shows $\Delta\phi_{14}$ increasing slightly
towards the edge of the beam, this is expected to occur when
$A_p(x)\rightarrow 0$, which degrades any accurate calculation of the
phase as the {\it signal--to--numerical noise} ratio decreases. At the beam
peak, shown in the right plot, the phase change is
$\Delta\phi_p(x=0)\approx20\mu$rad and
shows no correlation with $\delta x$. This result is 5 orders of magnitude smaller than what Eq.(\ref{eq:phaseshift})
states for displaced grating structures.
To determine whether the oscillations seen in
$\Delta\phi_p(x)$ were near field effects or numerical artefacts the
value $\max\{\Delta\phi_p(x)\}$ was computed at increasing distances from the WGM for a displacement over one grating period
 $p=0\rightarrow 28$ at the centre of the beam ($x=-d/2\rightarrow d/2$). As seen in Fig.\ref{fig:dhpi_disp}, the near field
phase shifts from the initial imprint of the grating can be seen at
$y<3\mu$m which decays rapidly with distance, after which a flat noise is present. Fig.\ref{fig:dhpi_disp} shows 3 different FFT windowing functions agreeing at
$y<3\mu$m but possessing different noise floors, the lowest being
$\max\{\Delta\phi_p(x)\}\approx10^{-7}$rad using a {\it Blackman} FFT
window. Numerical errors present from the FDTD are not thought to be limiting, increased spatial and temporal resolution ($\Delta x\rightarrow\Delta x/2$) does not offer any improvement in the noise levels as seen in Fig.\ref{fig:dhpi_disp}. This suggests {\it spectral leakage} from the FFT is limiting the accuracy of
phase measurements and the oscillations present in $\Delta\phi_p(x)$
measured at $15\mu$m are purely numerical artefacts, similar results were seen at varying distances from the WGM.

This work presents the successful implementation of an FDTD simulation to analyse
displacement induced phase shifts in a reflected Gaussian beam from a
WGM. No such phase shifts were found 
within the precision limit of $\approx10^{-7}$rad set by numerical errors.
This lower limit is seven orders of magnitude lower than the phase noise
estimated for previously proposed layouts with diffraction gratings,
which raised concerns regarding the stability and alignment
\cite{Freise07} of such configurations. Therefore, the absence of this
phase shift for WGMs strengthens the argument for their usage in
future high-precision interferometry experiments. 

We acknowledge the Science and Technology Facilities Council (SFTC)
for financial support in the UK. D.F. was supported by the Deutsche
Forschungsgemeinschaft (DFG) within the Collaborative Research Centre
TR7. This document has been assigned the LIGO Laboratory Document number LIGO-P1300019.

\begin{thebibliography}{10}
\newcommand{\enquote}[1]{``#1''}

\bibitem{PhysRevLett.100.013601}
H.~M\"uller-Ebhardt, H.~Rehbein, R.~Schnabel, K.~Danzmann, and Y.~Chen, Phys.
  Rev. Lett. \textbf{100}, 013601 (2008)

\bibitem{PhysRevLett.93.250602}
K.~Numata, A.~Kemery, and J.~Camp, Phys. Rev. Lett. \textbf{93}, 250602 (2004)

\bibitem{0264-9381-19-5-305}
G.~M. Harry, A.~M. Gretarsson, P.~R. Saulson, S.~E. Kittelberger, S.~D. Penn,
  W.~J. Startin, S.~Rowan, M.~M. Fejer, D.~R.~M. Crooks, G.~Cagnoli, J.~Hough,
  and N.~Nakagawa, CQG \textbf{19}, 897 (2002)

\bibitem{Chelkowski09}
S.~Chelkowski, S.~Hild, and A.~Freise, Phys. Rev. D \textbf{79}, 122002 (2009)

\bibitem{Sun:98}
K.-X. Sun and R.~L. Byer, Opt. Lett. \textbf{23}, 567 (1998)

\bibitem{Bunkowski06b}
A.~{Bunkowski}, O.~{Burmeister}, D.~{Friedrich}, K.~{Danzmann}, and
  R.~{Schnabel}, CQG \textbf{23}, 7297 (2006)

\bibitem{Brueckner08}
F.~Br\"{u}ckner, T.~Clausnitzer, O.~Burmeister, D.~Friedrich, E.-B. Kley,
  K.~Danzmann, A.~T\"{u}nnermann, and R.~Schnabel, Opt. Lett. \textbf{33}, 264
  (2008)

\bibitem{Wise05}
S.~{Wise}, V.~{Quetschke}, A.~J. {Deshpande}, G.~{Mueller}, D.~H. {Reitze},
  D.~B. {Tanner}, B.~F. {Whiting}, Y.~{Chen}, A.~{T{\" u}nnermann}, E.~{Kley},
  and T.~{Clausnitzer}, Phys. Rev. Lett. \textbf{95}, 013901 (2005)

\bibitem{deepaper}
D.~Lodhia, D.~Brown, F.~Br\"uckner, and A.~Freise, in
  preparation (2013)


\bibitem{Freise07}
A.~{Freise}, A.~{Bunkowski}, and R.~{Schnabel}, New Journal of Physics
  \textbf{9}, 433 (2007)


\bibitem{Sharon:97}
A.~Sharon, D.~Rosenblatt, and A.~A. Friesem, J. Opt. Soc. Am. A \textbf{14},
  2985 (1997)


\bibitem{taflove:2005}
A.~Taflove and S.~C. Hagness, {\it Computational Electrodynamics: The
  Finite-Difference Time-Domain Method} 3rd ed. (Artech House
  Publishers, 2005)

\bibitem{springerlink:10.1186/1687-6180-2012-56}
P.~Sysel and P.~Rajmic, EURASIP Journal on Advances in Signal Processing
  \textbf{2012}, 1 (2012). 10.1186/1687-6180-2012-56

\end{thebibliography}
\end{document}